\documentclass[epj]{svjour}
\usepackage{graphicx,amssymb}

\usepackage[bookmarksnumbered=true,linkbordercolor={0.8 0.8 
0.8},citebordercolor={0.6 0.7 0.6}]{hyperref}

\def\makeheadbox{{%
\hbox to0pt{\vbox{\baselineskip=10dd\hrule
\hbox
to\hsize{\vrule\kern3pt\vbox{\kern3pt
\hbox{Eur. Phys. J. B {\bfseries 66}, 259-269 (2008)}
\hbox{DOI: 10.1140/epjb/e2008-00419-y}
\kern3pt}\hfil\kern3pt\vrule}
\hrule
}%
\hss}}}

\begin{document}

\title{Level-dynamic approach to the  excited spectra of the Jahn-Teller
model - kink-train lattice and 'glassy' quantum phase }

\titlerunning{Level dynamics approach to the Jahn-Teller model}

\authorrunning{E. Majern\'{\i}kov\'a, S. Shpyrko}

\author{ Eva Majern\'{\i}kov\'a\inst{1}\fnmsep \inst{2} \fnmsep \thanks{\email{eva.majernikova@savba.sk}}
 \and S. Shpyrko\inst{2} \fnmsep \inst{3} \fnmsep \thanks{\email{serge\_shp(at)yahoo.com}} }


\institute{Institute of Physics, Slovak Academy of Sciences,
D\'ubravsk\'a cesta, 84511  Bratislava, Slovak Republic 
\and
Department of Optics, Palack\'y University,
T\v r. 17. listopadu 50, 77207 Olomouc, Czech Republic 
\and
 Institute for Nuclear Research, Ukrainian Academy of Sciences,
pr. Nauki 47, Kiev, Ukraine }

\date{Received: 12 June 2008 / Received in final form: 30 September 2008 \\ Published online 20 November 2008}

\abstract{The dynamics of excited phonon spectra of the $E\otimes e$
Jahn-Teller (hereafter, JT) model mapped onto the generalized
Calogero-Moser (gCM) gas of pseudoparticles implies a complex
interplay between nonlinearity and fluctuations of quasiparticle
trajectories. A broad crossover appears in a pseudotime (interaction
strength) between the initial oscillator region  and the nonlinear
region of the kink-train lattice as a superlattice of the
kink-antikink gCM trajectories. The local nonlinear fluctuations,
nuclei (droplets) of the growing kink phase arise at the crossover,
forming a new intermediate droplet "glassy" phase  as a precursor of
the kink phase. The "glassy" phase is related to a broad maximum in
the entropy of the probability distributions of pseudoparticle
accelerations, or level curvatures. The kink-train lattice phase
with  multiple kink-antikink collisions is stabilised by long-range
correlations when approaching a semiclassical limit. A series of
bifurcations of nearest-level spacings were recognised as signatures
of  pre-chaotic behaviour at the quantum level in the kink phase.
Statistical characteristics can be seen to confirm the coexistence
within all of the spectra of both regularity and chaoticity to a
varying extent (nonuniversality). Regions are observed within which
one of the phases is dominant. 
}

\maketitle

\section{Introduction}
\label{intro}

The analysis of irregularities of quantum spectra is essentially the search
for fingerprints of chaotic behaviour at the quantum level. It is
especially delicate for models lacking a reasonable
semiclassical limit due to the presence of quantum tunneling or
nonadiabatic fluctuations \cite{Gutzwiller:1990}. It is equally delicate
for systems with a combination of regular and chaotic phase space \cite{Ketzmerick:2007}.
Previous studies have shown that nonadiabatic fluctuations act to stabilise the
ground state, causing the ground state potential to become highly nonlinear \cite{Majernikova:2003}.
In this paper we shall investigate the consequences of the interplay
of nonadiabatic fluctuations due to level correlations
 and  nonlinearity in {\it complex excited spectra} of the E$\otimes$e JT
electron-phonon model in the situation of increasing interaction
strength.

The JT model, with one control parameter of electron-phonon
interaction strength, can be mapped onto a fully integrable
classical many-body system.  This system is observed to experience repulsive long-range interactions
known as the generalized Calogero-Moser gas (gCM) of Coulomb
interacting pseudoparticles
\cite{Pechukas:1983,Yukawa:1985,Nakamura:1986,Nakamura:1993}. This method provides a
bridge to a statistical description of the system and thus appears to be a basis
for differentiating between regularity and chaos in some systems
\cite{Burgdorfer:1992,Ishio:1992,Nakamura:2003}. It is worth noting that the
statistical description applies only for models with appropriate
Gibbs measures.

 In this paper, the authors seek to numerically and analytically examine level dynamics
of levels as functions of the parameter  (electron-phonon
coupling as a pseudotime) in pseudospace. Probabilistic considerations enable
determination of a measure of stochasticity within this irregular and highly non-universal model, in the context of earlier analyses of the nearest level spacing probability distributions \cite{Majernikova:2006a}.
Besides the numerical analysis we shall
apply an alternative approach of level dynamics.
We shall present a related approximate analytical calculation
to shed light on spectral structures, the physical mechanisms of their formation and the crossover between them. One advantage of this approach is the ability to
classically represent quantum fluctuations and examine
nonlinearity of the originally quantum system.

Excited energy levels $E_n(\alpha ),\ n=1,\dots , N$ of a quantum
Hamiltonian in the form $H(\alpha)=H_0+\alpha V$ can be considered
as dynamic coordinates of $N$ interacting classical pseudo-particles moving in
a pseudo-time $\tau\equiv \alpha$, $E_n(\alpha)\equiv x_n(\tau)$. The gCM set of equations
equivalent to the quantum mechanical problem
$H(\alpha)|\phi\rangle=E_n (\alpha)|\phi\rangle$ reads
\cite{Pechukas:1983,Yukawa:1985}
\begin{eqnarray}
\mathrm{d}x_n/\mathrm{d}\tau= p_n(\tau) \,, \qquad
\frac{\mathrm{d}p_n}{\mathrm{d}\tau}=2\sum_{m(\neq
n)}\frac{L_{nm}L_{mn}}{(x_m-x_n)^3}, \nonumber\\
\frac{\mathrm{d}L_{mn}}{\mathrm{d}\tau}=\sum_{l\neq
(m,n)}L_{ml}L_{ln}\left[\frac{1}{(x_n-x_l)^2}-\frac{1}{(x_m-x_l)^2}\right
], \label{C1}
\end{eqnarray}
where
\begin{eqnarray}
\lefteqn{L_{mn}(\tau) =\Big(x_n(\tau)- x_m(\tau)\Big)\cdot V_{mn}(\tau)=-L_{nm} \,, \quad} \hspace{-1ex} \label{C1'}
\\
&&p_n \equiv \big \langle n(\tau)|V|n(\tau)\big \rangle=V_{nn}(\tau) , \,
V_{nm}(\tau) = \big\langle n(\tau)|V|m(\tau)\big\rangle.\nonumber
\end{eqnarray}
This system is known to be completely integrable
\cite{Nakamura:1986,Nakamura:1993}. Hence, it possesses as many
integrals of motion as it has independent variables. The
most important two additive integrals of motion are the total
"energy" (the classical Hamiltonian which generates the above
equations of motion (\ref{C1}))
\begin{eqnarray}
E=
\frac{1}{2}
\sum_{n=1}^N p_n^2-\sum _{1\leq j < k\leq N}V_{jk}^2(\tau)\log
|x_j-x_k| \label{C1''}
\end{eqnarray}
and the total "angular momentum" \cite{Yukawa:1985}
\begin{equation}
Q=\sum _{1\leq j < k<\leq N}L_{jk}^2\,. \label{C1'''}
\end{equation}
The first or both integrals of motion were used to develop the
statistical description of the gCM system by an appropriate choice of
canonical ensemble and corresponding Gibbs measures
\cite{Yukawa:1985,Gaspard:1990}.

Equation (\ref{C1''})  allows us to understand the system as a
 two-dimensional Coulomb plasma of non-local time dependent
pseudocharges $V_{jk}(\tau)$. The respective set of equations
alternative to the equations for $L_{mn}$  is implied by (\ref{C1})
and (\ref{C1'}) to yield \cite{Burgdorfer:1992}
\begin{eqnarray}
\frac{\mathrm{d}V_{mn}}{\mathrm{d}\tau} & = & \sum_{l\neq
(m,n)}V_{ml}(\tau)V_{ln}(\tau)\left[\frac{1}{x_m-x_l}+\frac{1}{x_n-x_l}\right
]\nonumber\\
& & -V_{mn}(\tau)\frac{p_m-p_n}{x_m-x_n} \,.  \label{Vmn}
\end{eqnarray}
 The formalism of level dynamics (\ref{C1}-\ref{Vmn}) may be applied to the
excited phonon spectrum of the E$\otimes$e JT
Hamiltonian \cite{Majernikova:2003,Majernikova:2006a}

\begin{equation} H=  (b_{1}^{\dag}b_{1} +b_{2}^{\dag}b_{2}+1 )I +
\alpha (b_{1}^{\dag}+b_{1})\sigma_{z}
 -\alpha (b_{2}^{\dag}+b_{2})\sigma_{x}.
 \label{1}
\end{equation}
In this equation, $b_{1,2}$ represent boson (phonon) operators of two oscillators with
the frequency $\Omega=1$, $\sigma_x=\left (\matrix {0 & 1\cr 1 &
0}\right )$, $\sigma_y=i\left (\matrix{ 0 & -1   \cr 1 &  0  } \right )$,
$\sigma_z=\left (\matrix {1 & 0 \cr 0 &  -1}\right )$ are Pauli
matrices, $I$ is the unit matrix, $\alpha$ is the coupling constant
between the electron and phonon modes. The $2\times 2$ matrix form
accounts for two electron levels.
 According to the above, the model (\ref{1}) with one non-integrability parameter
 $\alpha$  ($H(\alpha)=H_0+\alpha V$) and the integrable part $H_0$ is suitable
for gCM mapping (\ref{C1}) to be applied. It is worth noting that the
case containing different frequencies of the phonon modes $1$ and $2$ can easily be
transformed into a case with two different coupling constants
$\alpha\neq\beta$ in the two last terms of (\ref{1}). This was investigated in a series of previous papers
\cite{Majernikova:2003,Majernikova:2006b,Majernikova:2008}. However,
the level dynamics approach for systems with several
nonintegrability parameters is much more complicated
\cite{Nakamura:1993,Nakamura:1992}. Analysis reveals
non-trivial gauge structures induced by multidimensional
parameter space.

The JT model (\ref{1}) consists of two degenerate electron levels
coupled with two vibron modes of different symmetry against
reflection. It is a typical representation of a nonintegrable
nonadiabatic system posessing no reasonable semiclassical limit. The
model has been intensively studied over time
\cite{Majernikova:2003,Nakamura:2003,Long:1958,OBrien:1964} and is
known to be a rich source of understanding for physical properties and
consequences, for both
heuristic interest and practical applications. The rotationally symmetric
E$\otimes$e model, besides of the common $SU(2)$ reflection
symmetry,  has one additional constant of motion -- the
conserved angular momentum
 $\hat{J}=i(b_1 b_2^+-b_1^+b_2)-\frac{1}{2}\sigma_y$ with eigenvalues $j=1/2,3/2,\dots$\,.
 Hence, in the following, the representation of the definite quantum number $j$
will be used which appears to be an
 additional parameter of the model \cite{Majernikova:2006a}.
 To illustrate the typical properties of the gCM gas
 we have numerically solved the set of gCM dynamical
 equations (\ref{C1}), (\ref{C1'}) and (\ref{Vmn}) for the system
 (\ref{1}).
The initial conditions $V_{mn}(0)$ for the
model (\ref{1}) are specified from the set of equations for the excited
levels of the model. It consists of pairs
of even and odd levels $E_n$, $n \equiv 2n_r,\  2n_{r}+1 $, $n_r=0,1,\dots$
(main quantum number)
determined by the well known tridiagonal matrix equations for energy $E$

\begin{eqnarray}
&(E_{2n_r}^0-E)c_{2n_r} + \alpha
(f_{n_r,n_r}c_{2n_r+1}+f_{n_r,n_{r-1}}c_{2n_r-1})=0  , \quad \label{cc}
\\
& (E_{2n_r+1}^0-E )c_{2n_r+1} + \alpha
(f_{n_r,n_r}c_{2n_r}
+f_{n_{r+1},n_r}c_{2n_r+2})=0\,. \nonumber
\end{eqnarray}
In this equation,  $E_{n}^0$ represent the energies of unperturbed harmonic
oscillators; $\{c_n\}$ are components of the wave function in the
harmonic oscillator representation ($H_0$ is diagonal). The
perturbations  terms in  (\ref{cc}) are represented by the matrix
elements \cite{Majernikova:2006a,Long:1958}
\begin{eqnarray}
V_{2 n+1, 2n}(0)\equiv & f_{n_r n_r} & =  \sqrt 2 \sqrt{n_r +1+|j-1/2|}, \nonumber \\
 V_{2n-1, 2n}(0)\equiv & f_{n_r n_r-1} & =  -\sqrt 2 \sqrt{n_r} \,,
\label{V:init}
\end{eqnarray}
giving  the values of pseudocharges (\ref{Vmn}) at $\tau=0$.


\section{GCM dynamics of the Jahn-Teller excited
spectra. Numerical results}
\label{num}

\begin{figure}[ht]
\includegraphics[scale=0.5]{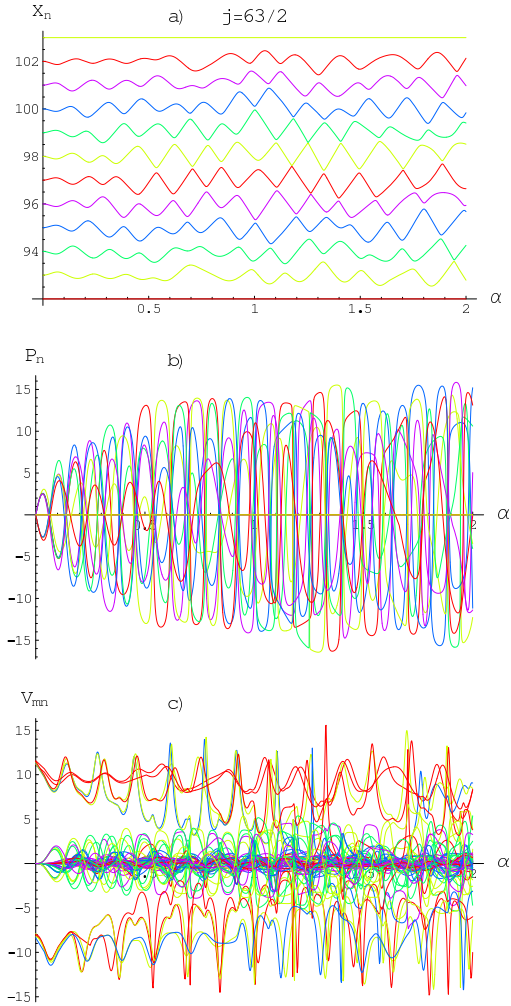}
\caption{ Numerical solution of level
 dynamics equations (\ref{C1})-(\ref{C1'}) in the $ n_r>j$ part of the Jahn-Teller
 excited spectra for
$j=63/2$ with artificial boundary conditions (see the beginning of Sect. 2).
The levels from $n=93$ to $n=102$ are
shown provided that those at $n=92$ and $n=103$ are kept fixed. (a), (b)
show respectively ``coordinates'' $x_n$, and related momenta
$p_n\equiv V_{nn}$, and (c) shows all the off-diagonal matrix
elements $V_{mn}$ of the perturbation $\hat{V}$ for $m,n$ within the
described range. The transition between the oscillatory (small $\alpha$)
and the kink region with avoidings ($p_n=0$) is marked by mixing of
the $V_{mn}$ branches. The kink lattice region is characterised by strong mixing of three $V_{mn}$ branches. Narrow windows of
weaker mixing at large $\alpha$ are apparent.} \label{fig1}
\end{figure}
\begin{figure}[htb]
\includegraphics[scale=0.55]{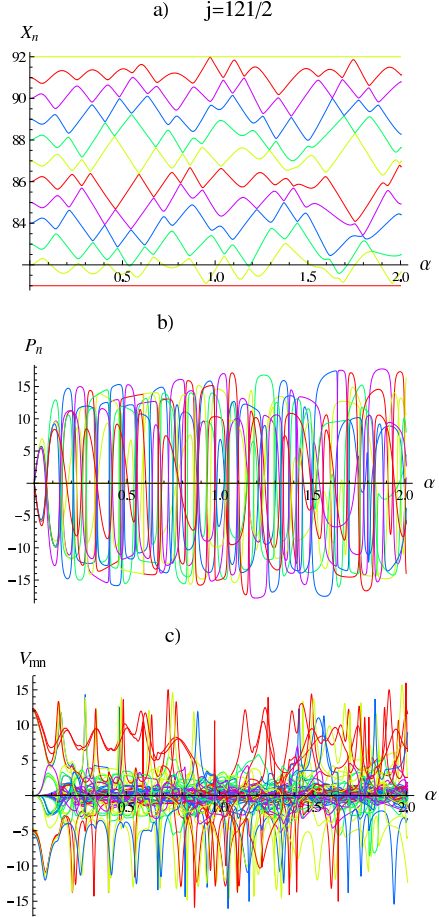}
\caption{The same for $j=121/2$, $ n_r <j$. Note the kink-lattice
region $x_n \sim v\alpha $ at strong mixing of three $V_{nm}$
branches. Narrow segments of collisionless
ballistic dynamics appear with linear time-dependent trajectories $x_n$ and
constant momenta $p_n$ at $V_{mn}=0$. With increasing $\alpha$ the
"oscillating" branches are absorbed by the central stochastic band.
} \label{fig2}
\end{figure}

In this section we present a numerical solution for the set of gCM
equations (\ref{C1}-\ref{C1'}) for several values of rotational
quantum number $j$. The initial conditions $x_n(0)=E_n^{(0)}$,
$p_n(0)=0$, $V_{2 n+1, 2n}(0)$, $V_{2n-1, 2n}(0)$ are given by
(\ref{V:init}). Here and in subsequent sections the
index $n$ labels the energy levels in increasing order while $n_r$
and $j$ represent the quantum numbers of the initial unperturbed
oscillators \cite{Majernikova:2006a}.
  Typically the reduced sets of equations for twenty levels were examined.
  In order to confine the levels we imposed boundary conditions rendering
   the first and last levels unmovable, in other words, replacing the first and
last pseudo-particle by the pseudo-particles of infinite mass.
  The examples of numerical solutions to gCM equations are shown in
Figures \ref{fig1}, \ref{fig2} for
  three sets of variables ($x_n$, $p_n$, $V_{mn}$) involved.
  The solutions for energy levels not close to artificial interval boundaries
  locally effectively reproduce the numerical solutions for energy
levels as functions
  of $\alpha$ obtained via the diagonalisation of the Hamiltonian matrix (\ref{cc})
  (see \cite{Majernikova:2006a}, Fig.1). There the exact level spectrum is
illustrated for two different domains introduced
as ``oscillating'' and ``kink-lattice'' domains of the spectrum. These
 roughly correspond to the different types of behaviour shown in
Figures \ref{fig1}, \ref{fig2}. As the number of levels increases, so
does the effectiveness of the solutions at reproducing numerical
solutions via diagonalisation of the Hamiltonian.

\begin{figure}[hb]
\includegraphics[width=0.55\hsize]{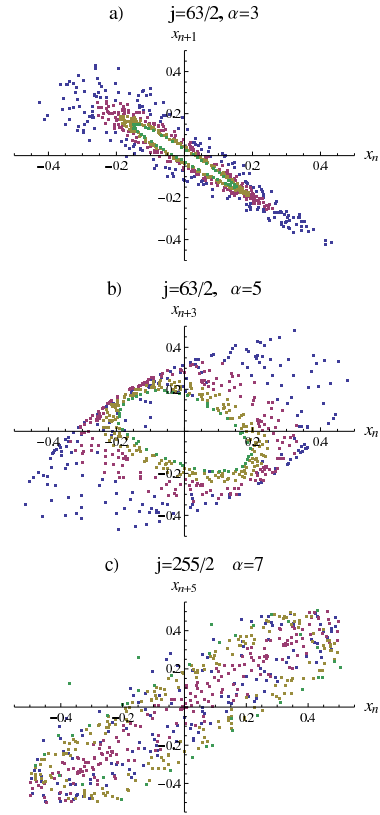}
\caption{Examples of maps of exact solutions $x_{n+i}$ vs $x_n$ for
$i=1,3,5$ for different $j$ and $\alpha$ and different parts of the
spectra: levels with numbers 100-300 - blue, 300-500 - red, 500-700
 - yellow, 700-750 - green. }
\label{fig3}
\end{figure}

 In Figures \ref{fig1}, \ref{fig2} one can recognise a complex
dynamical behaviour of three coupled groups of excitations: at small
$\alpha$ there are two groups of pseudo-particles (levels), even and
odd, of weakly perturbed oscillations with opposite amplitudes (Fig.
\ref{fig1}a) and  related rapidly oscillating pseudo-momenta
$p_n$ (Fig. \ref{fig1}b).
 After a small initial period
the level degenerations are removed due to the creation of new
inter-level correlations of more distant levels represented by the
"pseudocharges" $V_{mn}(\tau )$ for $m> n+1$ (Eqs. (\ref{C1},
\ref{C1'})). They emerge as the third (central) band in Figures
\ref{fig1}, \ref{fig2},c. With increasing $\alpha $ and $j$ the
interference between the central and side bands
 increases, destroying the oscillatory behaviour of the
levels $x_n$ and related momenta $p_n$. This scenario tends to the
formation of a single stochastic central band of pseudocharges with
increasing pseudotime (Fig.\ref{fig2}c). Moreover, within the single
stochastic band appear very narrow windows of collision-free
ballistic  motion, with linear dependence on $x_n\sim v \alpha $ and
$p_n\sim {\rm const}$. This may be attributable to the kink domain in
 the long time limit  when the joint effect of the stochastic set of
correlations (including the long-distance inter-level correlations
(Fig. \ref{fig3}c)) vanishes, allowing a situation similar to free
particle dynamics. We
shall comment further on the structure of the spectra on the basis of the
analytical results contained in Section \ref{analyt}.

 Correlations between the levels  ($x_{n+i}$, $x_n$),
$i=1,3,5$ (maps) for different model parameters and parts of the
spectra are illustrated  in Figure \ref{fig3}. In these figures the
notation $x_n$ stands for the ``reduced'' part of the level energy
with extracted energy of the level $E_0$ and ``secular'' part
$n-\alpha^2$ (Cf. Fig. 4 of \cite{Majernikova:2006a}).
 The semi-elliptic shape of the maps, related to the highest parts of the
spectra, suggests that the correlations $x_{2n+1}\cdot x_{2n}$ have a strong effect
 with excluded space inside the ellipse (Figs.
\ref{fig3}a, \ref{fig3}b). This is a direct indication of level avoidings. The
effect noticeably changes for different parts of the spectra and
becomes more pronounced and unpredictable with increasing $\alpha$ and
$j$. On the other hand, for extremely high $\alpha$ and $j$ close to
the classical limit, the effect of correlations randomises the map
so strongly that the excluded space substantially shrinks to several
regions of a possible fractal structure induced by
 the long-range correlations (Fig. \ref{fig3}c).

\subsection{Probabilistic distributions of pseudo-particle characteristics}

\begin{figure}[hb]
\includegraphics[width=0.55\hsize]{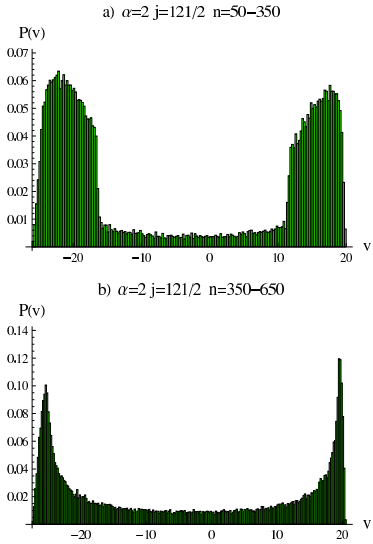}
\caption{Probability distributions of velocities of pseudoparticles
$P(v)$, $v\equiv V_{mm}(\tau)$. Peaked oscillator-dominated
region in the upper part of the spectra (b);
Stochasticity-dominated broad bands in the lower part of the spectra
(a).} \label{fig4}
\end{figure}

A complementary approach to the developed dynamic (deterministic)
picture can be provided by describing the system in terms of
statistical distributions. The common {\it ad hoc} assumption
is to introduce a canonical (or grand canonical)
ensemble and an appropriate Gibbs measure in the phase space of all
dynamical variables \cite{Yukawa:1985,Gaspard:1990} $\mathrm{d}M=\exp(-\beta
H) \mathrm{d}x_n\mathrm{d}p_n\mathrm{d}L_{mn}$. In the domain of
developed quantum chaos, this
approach was shown to reproduce the main results of random
matrix theory (RMT). In particular, it is effective in predicting the form of the
distribution of level spacings and level velocities. Similarly, it
can be used for the investigation of the response of the energy
spectrum to the change of nonintegrability parameter (that is for
the distribution of pseudoparticle  velocities $v\equiv p_n= \mathrm{d}
x_n(\tau)/\mathrm{d}\tau = V_{nn}(\tau )$ and their accelerations (level
curvatures) $P(K)$, $K\equiv \mathrm{d}^2x_n(\tau)/\mathrm{d}\tau^2$). In particular,
in the domain of applicability of the RMT the distribution of the
velocities should be Gaussian with the dispersion equal to the
``temperature'' parameter $1/\beta$ in the Gibbs measure
\cite{Burgdorfer:1992}. The principal prediction for the
distribution of the level curvatures is their asymptotic behaviour
at large $K$ which has been shown universally to be of the form $P(K) \sim
K^{-\nu}$ with $\nu=3,4,6$ for different kinds of ensembles
\cite{Gaspard:1990}. The strongest assumption made in this approach is
determining the form of the Gibbs measure, which implies that the system
is in thermodynamic equilibrium. In the domains of undeveloped chaotic behaviour
this assumption appears invalid. In
particular, the statistical distributions $P(v)$ of the level
velocities $v\equiv p_n\equiv V_{nn}(\tau)$ for fixed $j$ and
$\tau\equiv \alpha$ for JT system are shown in Figure \ref{fig4}.
 They develop
from an initial (at small $\alpha$) oscillation for large
$n$ (upper part of the spectra, Fig. \ref{fig1}c, \ref{fig4}b) to strongly broadened
oscillatory bands due to the stochastic velocities in the lower part
of the spectra, Figure \ref{fig2}c, \ref{fig4}a. Two distinct bands of positive and
negative velocities and the in-gap  remnants of the oscillations
seen in Figure \ref{fig4}a are interpreted in Section \ref{analyt}.

\begin{figure}[hb]
\includegraphics[width=0.6\hsize]{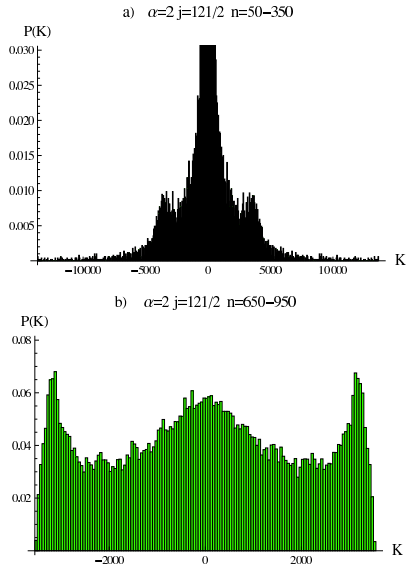}
\caption{Probability distributions  $P(K)$ for $\alpha=2$,
$j=121/2$, where $K=\partial^2 x_n/\partial \tau ^2$ is the level
curvature (pseudoparticle acceleration). Note the two-peaked
oscillator-dominated region in the upper part of the spectra (b).
The stochasticity-dominated region in the lower part of the spectra
exhibits (i) a prominent peak at $K=0$ which corresponds to the
kinks ($p=const$, see Fig. \ref{fig2}b); (ii) characteristic for RMT
prediction of asymptotic power-like dependence at large $K$. The
side "wings" of the distributions are in contrast to the predictions
of the RMT and indicate a presence of memory terms in the equation
for the measure $\mathrm{d} M (\alpha)$. } \label{fig5}
\end{figure}

 The non-Gaussian
character of the distributions $P(v)$ in Figure \ref{fig4} and  $P(K)$, Figure
\ref{fig5} exposes the necessity of more correctly introducing the
stochastic measure. As the authors have previously suggested
\cite{Majernikova:2006a}, the development of the distribution of
$P(K)$ in ``time'' $\alpha$ has some features common with the
diffusion process governed by the diffusion equation with additional
telegraph term, i.e. $T \partial^2 P /\partial \tau^2$ $+$ $\partial P
/\partial \tau$ $=$ $D
\partial^2 P/\partial K^2$, where $D$ is the diffusion
coefficient. The coefficient of the telegraph term accounts for
memory effects. In particular, it is responsible for the
appearance of the characteristic ``wings'' for large $K$ (Fig. \ref{fig5}b)
which are not observed through the RMT approach. A correct explanation for
these features requires introducing an equation for which the measure
$\mathrm{d}M(\alpha)$ developes in time and only coincided with the Gibbs measure
in the limit of RMT (``quantum chaos'').  In view of
the two-peak character of the probability distributions $P(K)$
we use their ``entropies'' $S_K=-\int P(K) \log P(K) \mathrm{d}K$
to characterize their properties rather than
their dispersion characteristics \cite{Majernikova:epjb:2004}.

\begin{figure*}[htb]
\includegraphics[width=0.6\hsize]{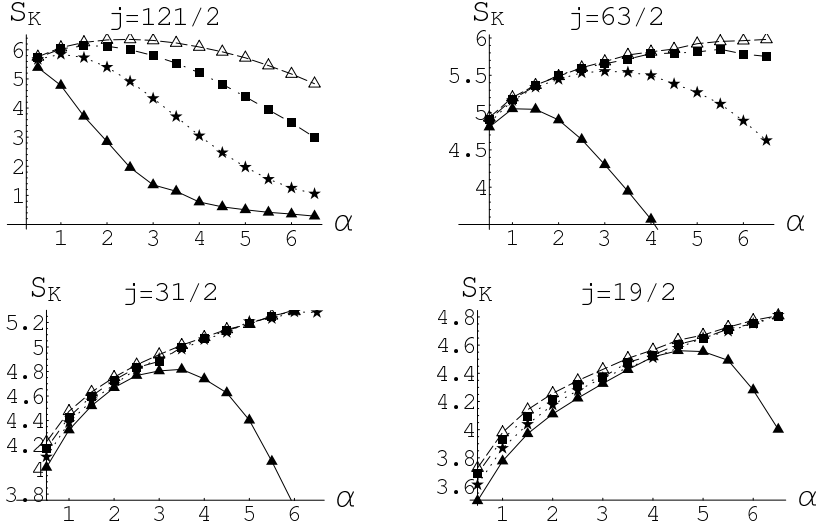}
\caption{Entropies $S_K$ of the probability distributions of the
accelerations (curvatures) $P(K)$ averaged over successive
intervals of 100 levels. The maximum irregularity
(stochasticity) appears in the region with maximum mixing of three
branches of the pseudocharges identified as the region of droplet
formation and as a precursor phase of the kink-train lattice. Triangles -
levels 50-150; stars: levels 150-250;
squares: 250-350; open triangles: 350-450. The entropies are given
with an arbitrary constant of the form $\Delta K \log \Delta K $
accounting for the finite sampling bins $\Delta K$ on the histograms
such as shown in Figure \ref{fig5}, which were the same ($\Delta K=10$) for all
graphs. }
 \label{fig6}
\end{figure*}

The entropies $S_K$ of the probability distributions of the
accelerations (curvatures) $P(K)$ averaged over successive intervals
of 100 levels are plotted in Figure \ref{fig6} as functions of pseudotime $\alpha$.
The crossover between the two above described phases is evident. These figures show
the broad maximum irregularity (stochasticity) in the region of the maximal
mixing of three branches of the pseudocharges identified as the
region of growing nonlinear fluctuations (formation of droplets) as
a precursor phase of the kink-train lattice (see subsequent section).

\vspace{-2em}

\section{GCM dynamics of the Jahn-Teller excited
spectra. Approximate analytical approach} \label{analyt}

The simplest analytical approach based on the approximation of
constant (initial) values of matrix elements $V_{2n+1,2n}(\tau)$ and
$V_{2n-1,2n}(\tau)$ can be applied for a small $\alpha$. It must be
borne in mind that this approach neglects the dynamics of
$V_{mn}(\tau)$, the correlations of more distant levels
developed in the course of pseudotime (Figs.\ref{fig3},b,c).  As a
consequence  our two-level approach includes only the dominant
nearest neighbour interactions. To a limited extent this
approximation allows for understanding the main features of the
model and is satisfactorily consistent with our numerical results. This
formulation recovers the local crossover between the anharmonic
oscillator region and the complex nonlinear one.

Within the approximation of constant  matrix elements in (\ref{C1'}) it may be taken that
  $V_{2 n+1, 2n}(\tau )\approx V_{2 n+1, 2n}(0 )= f_{nn}(n_r,j)$,
  $V_{2n-1, 2n}(\tau)\approx  V_{2n-1, 2n}(0)= f_{n,n-1}(n_r,j)$ from (\ref{V:init})
(in what follows we denote $j\equiv |j-1/2|$ for brevity).
  Then,
\begin{eqnarray}
 L_{2n+1,2n}(\tau) & = & V_{2 n+1, 2n}(0)\big(x_{2n+1}(\tau)-x_{2n}(\tau)\big)\,, \nonumber\\
L_{2n-1,2n}(\tau) & = & V_{2n-1, 2n}(0)\big(x_{2n}(\tau)-x_{2n-1}(\tau)\big) \,.
\label{C2}
\end{eqnarray}
Since  the number of phonons is not conserved  the number of
pseudoparticles (excited levels) in the system with Hamiltonian
(\ref{1}) is not constant. Therefore
a two-level approximation of the gCM equations can be used, but only
for small $\alpha$. Denote the corresponding levels as $x_1$, $x_2$.
The origin may be set so that two levels $x_1, \ x_2$ are
symmetric and separated by $x_2-x_1\equiv 2 x$; the effect of the
other levels of the system can be accounted for by setting two bounding (fixed)
levels, $x_0, \ x_3$ separated by a distance $x_3-x_0=3$. To
simplify notations we denote $f_{nn}\equiv  f_0, \ f_{n+1,n}\equiv
f_1$. The set of gCM equations then yields

\begin{eqnarray}
\left\{
\begin{array}{l} {\displaystyle
\frac{1}{2}\frac{\partial p_1}{\partial \tau}=
\frac{-f_{0}^2}{x_2-x_1}+
\frac{f_{1}^2}{x_1-x_0} } \\ {\displaystyle
\frac{1}{2}\frac{\partial p_2}{\partial
\tau}=\frac{f_{0}^2}{x_2-x_1}- \frac{f_{1}^2}{x_3-x_2} \,.}
\end{array}
\right.
\label{2-0}
\end{eqnarray}

\begin{figure}[hb]
\includegraphics[width=0.7\hsize]{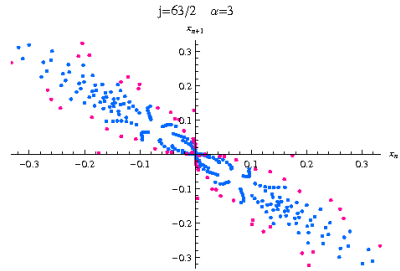}
\caption{Mapping of approximate solutions $x_{n+1}$ vs $x_n$ (as in Fig.\ref{fig3}a) generated by the simple two-level
model (\ref{2-0}). The correspondence with the exact result in
Figure \ref{fig3} is satisfactory except in the near-zero part
since the level avoidings are neglected in this model. } \label{fig7}
\end{figure}

The equation for the separation $x$ of the two nearest
trajectories reads
\begin{equation}
\ddot x =-\frac{\partial V(x)}{\partial x}=\frac{2f_0^2}{2x} -
\frac{2f_1^2}{3/2-x}, \label{linear}
\end{equation}
where $V(x)= -f_0^2\log x -2f_1^2 \log (1.5-x)$ is the
one-dimensional time-dependent Coulomb potential with
 $x_{eq}= 3 f_0^2/2\left(f_0^2+2f_1^2\right)$ as a
point of the potential minimum where $V^{'}(x_{eq})=0$. \\
The harmonic oscillations of the frequency $V''= 2
(2f_1^2+f_0^2)^3/(3f_0f_1)^2$ become unstable at $V^{'''}=0$, or when
$f_0^2=2f_1^2$ or $n_r\approx j$. This condition determines the
border between the harmonic oscillation region ($n_r>j$) and the
nonlinear region ($n_r<j$) in this simplified model. The model
generates the map illustrated in Figure \ref{fig7}.  Since the
subsequent clusters in it are independent, the whole map is
completely determined by the phase shift of the oscillations from
cluster to cluster (changing the parameter $n_r$ numbering the
clusters). Resemblance to the corresponding exact results, Figure \ref{fig3}a
can be traced only for higher levels otherwise the simple two-level
symmetric model completely ignores the level avoidings.

In order to proceed beyond the two-level harmonic approximation we shall
 account for the collective effect of neighbour levels with
level-to-level variations of the ``pseudocharges'' $f_{mn}$.  Let us
define small fluctuations $\delta $ as $x_{2n+1}-x_{2n}\equiv
1+\delta_{2n}$. The set of equations derived from
(\ref{cc}) then reads
\begin{eqnarray}
\frac{1}{2}\frac{\partial^2\delta_{2n}}{\partial
\tau^2} &=& \frac{2f_{nn}^2}{1+\delta_{2n}}-\frac{f_{n+1,n}^2}{1+\delta_{2n+1}}-
\frac{f_{n,n-1}^2}{1+\delta_{2n-1}}
\nonumber\\
\frac{1}{2}\frac{\partial^2\delta_{2n+1}}{\partial
\tau^2}  &=&  \frac{2f_{n+1,n}^2}{1+\delta_{2n+1}}-\frac{f_{n+1,n+1}^2}{1+
\delta_{2n+2}}-\frac{f_{nn}^2}{1+\delta_{2n}}.
\label{pairs}
\end{eqnarray}
 For sufficiently large $n$, $f_{n+1,n+1}\approx f_{nn},\
f_{n+1,n}\approx f_{n,n-1}$, hence equations (\ref{pairs}) imply
the approximate relations $\delta_{2n+1}$ $\approx \delta_{2n-1}$ and
$\delta_{2n+2}\approx \delta_{2n}$. Then the pairs of
pseudoparticles (levels) represent an analogy of a dimerized system
or a lattice with two atoms in a cell. It is worth noting that
from (\ref{pairs}) it follows that $\delta_{2n},
\delta_{2n+1}$ is a pair of reflection-symmetric variables against
the transformation $\tau\rightarrow i\tau $.

\subsection{Structure of the nonlinear  region and quantum phase
transition-like nature of the crossover}
\label{nonlin}

For increasing $\tau$ and $j$ the picture of ``pseudocharges''
$V_{mn}(\tau)$ (Eq. (\ref{C1''}) and Fig.\ref{fig1})
dramatically changes by the creation of new long-range components
which form a central branch of highly irregular oscillations. A
strong mixing between the short-range (two side-bands in
Fig.\ref{fig1}) and long-range $V_{mn}$ (central band) proceeds
towards a limit of large j, when the side-bands disappear
(Fig.\ref{fig2}c). At large $\alpha$, the level dynamics is mostly affected
by the central irregular branches of the ``pseudocharges''
$V_{mn}$. A simple approximate dynamic description of the crossover
between oscillatory and nonlinear regions can be obtained from the dynamic equations (\ref{pairs}) for the
fluctuations, $\delta_{2n}$ and $\delta_{2n+1}$.  We develop the
terms on r.h.s. of equations (\ref{pairs}) into a series up to the
lowest order of nonlinearity $\delta^3$. In the following
approximate calculations we shall move to a continuum
approximation in the coordinate $n$ labeling the level system. If we
extract the term
\begin{equation}
\delta_{2n+2}+\delta_{2n}-2\delta_{2n+1}\rightarrow \frac{\partial^2
\delta_{2n+1}}{\partial n^2 } \equiv \delta_{2n+1}^{''}
\label{deriv}
\end{equation}
and define $\delta_{2n+1}\equiv 1+y, \
\delta_{2n}\equiv x$, the equation for $\delta_{2n+1}$ of (\ref{pairs})
can be rewritten in  a  suitable form for further analysis,
\begin{eqnarray}
\lefteqn{\frac{1}{2} \ddot{y} -f_{n+1,n+1}^2  { y}^{''}=
f_{n+1,n+1}^2-f_{nn}^2+} \hspace{-1em} \nonumber\\
& & 2\left(f_{n+1,n+1}^2- 2f_{n+1,n}^2\right) y
-2f_{n+1,n}^2 y^2 ( y+2) \label{1otr} \\
& & -\left(f_{n+1,n+1}^2-f_{nn}^2\right)x- \left(f_{n+1,n+1}^2+f_{nn}^2\right) (x^2-x^3), \quad
\, \nonumber
\end{eqnarray}
 where the index $n$ of $y_n$ has been omitted for simplicity.
 Using a periodicity condition  mod$(y+2)= y$ the equation (\ref{1otr}) is rewritten as follows:
\begin{equation}
\frac{1}{2} \ddot{ y} -f_{n+1n+1}^2  { y}^{''}=-\frac{\partial V(
y)}{\partial  y} \, , \label{phasetr}
\end{equation}

where the nonlinear potential
\begin{eqnarray}
\lefteqn{ V( y)= \frac{1}{2} f_{n+1n}^2 y^4-
\left(f_{n+1,n+1}^2-2f_{n+1,n}^2\right) y^2-} \hspace{-1em} \label{Vy} \\
& & \left(f_{n+1,n+1}^2-f_{nn}^2\right)(1-x) y
+\left(f_{n+1,n+1}^2+f_{nn}^2\right) x^2\left(1-x\right) y
 \nonumber
\end{eqnarray}
indicates a crossover analogous to the  quantum first order phase
transition due to a small linear driving force $\sim y$, ($|x|\ll
|y|$ by definition). With the simplification valid for large $n$,
$f_{nn}^2\approx f_{n+1 n+1}^2, \ f_{n+1n}^2\approx f_{nn-1}^2$ one can
obtain a condition defining the dimerisation region
$\delta_{2n}\approx - \delta_{2n-1}$ where equation (\ref{pairs})
for $\delta_{2n}$ yields an equation analogous to (\ref{phasetr}).
 In the case when the driving forces on the r.h.s. of (\ref{Vy}) ($\sim y$) are
 either mutually compensated or both tend to zero (large $j$ and $n_r$),
that is,
 $\left(f_{n+1,n+1}^2-f_{nn}^2\right)(1-x)\sim
\left(f_{n+1,n+1}^2+f_{nn}^2\right) x^2(1-x)\ll f_{nn}^2$,
 equations (\ref{phasetr}) and
(\ref{Vy}) imply a crossover analogous to the quantum second order
phase transition at $f_{n+1,n+1}^2-2f_{n+1,n}^2 = 2(j-n_r) =0$. This
coincides with the criterion for the crossover to nonlinear
 region from simple analysis of equation (\ref{linear}).  For
$j<n_r$, $ f_{n+1,n+1}^2-2f_{n+1,n}^2 <0$,  $V(y)={\displaystyle \frac{1}{2}}
f_{n+1,n}^2 y^4- (f_{n+1,n+1}^2-2f_{n+1,n}^2) y^2$ becomes a
single-well potential, and we identify the phase of anharmonic
oscillators. At $j\geq n_r$ one has $f_{n+1,n+1}^2-2f_{n+1,n}^2
>0$. Here a double potential well of $V(y)$ opens, and correspondingly
a new phase appears. Equation (\ref{phasetr}) with the
driving force omitted acts as an analogy to a quantum second order phase
transition for each level (trajectory) $n$. However, the small
driving force in (\ref{Vy}) changes this scenario before it vanishes
at large $j$ and $\alpha\equiv \tau$. Examining equation
(\ref{1otr}) with accounting for a constant ($y$-independent)
driving force
\begin{equation}
F= \left[-(f_{n+1,n+1}^2-f_{nn}^2)+(f_{n+1,n+1}^2+f_{nn}^2)x^2
\right](1-x)>0 \label{F}
\end{equation}
 results in
\begin{equation}
\frac{1}{2} \ddot{ y} -f_{n+1,n+1}^2  {y}^{''}-
2(f_{n+1,n+1}^2-2f_{n+1,n}^2) y +2f_{n+1,n}^2  y^3 =-F \,, \label{force}
\end{equation}
here, $F$ in equation (\ref{F}) consists of two competing parts: the drift term
due to the broken periodicity of the level spectra at finite $\tau $
and the ``pressure'' force from the neighbouring levels. If $F>0$ and
$f_{n+1,n+1}^2-2f_{n+1,n}^2>0$ it drives a first order quantum phase
transition-like crossover generating nonlinear fluctuations.
The equation of type (\ref{force}) in the dimension $d>1$ has been
investigated  \cite{Langer:1967}  as a prototype for a
 first order phase transition involving the coexistence of two phases, one phase represented
 by  growing droplets (bubbles) spanned by the force $F$ in the sea of the other competing
 phase.
 In our case the dimension $d=1$; Nevertheless, there exists an exact solution to the
 normalized nonlinear
equation of the type (\ref{force}). It was found in the closed form of
a nonlinear oscillation \cite{Lal:1986} which in our case reads
\begin{equation}
 y \left(w\left(2\tau-\frac{n_r}{\sqrt j}\right)\right)=
a\frac{n_1+\cos \left(w\left(2\tau-{\displaystyle\frac{n_r}{\sqrt j}}\right)\right)}{n_2+\cos
\left(w\left(2\tau-{\displaystyle\frac{n_r}{\sqrt j}}\right)\right)}, \label{droplet}
\end{equation}
where $$n_1=\frac{2-n_2^2}{n_2},\ a^2=
\frac{f_{n+1,n+1}^2-2f_{n+1,n}^2}{f_{n+1,n}^2}\frac{n_2^2}{2+n_2^2}
>0\, ,$$
$$ w^2= \big(f_{n+1,n+1}^2-2f_{n+1,n}^2\big)\frac{n_2^2-1}{n_2^2+2}\, ,$$
$$a = \frac{F}{\displaystyle f_{n+1,n+1}^2-2f_{n+1,n}^2}\geq 0 \,.$$

 The ``droplet'' fluctuation (\ref{droplet}) is a traveling periodic non-sinusoidal
oscillation moving with velocity $v\approx 2\sqrt j$. The
amplitude $a$ of the fluctuation $ y$ is spanned by the driving
field $F$. This force grows when $ x$ ($x\leq x_{max}=1/\sqrt
{2(j+n_r)}$) decreases until the energy of the excitation
(\ref{droplet}) reaches the energy of the kink, where
$v\approx 2\sqrt j$ is the minimum velocity of the kink. If
$j$ increases, $j\gg n_r$, then $x\rightarrow 0$ and $a\rightarrow
0$ ($a\rightarrow 1/(j-n_r) $), so that the intermediate ``droplet''
 glassy phase transforms into the more stable kink lattice phase
(see below).

Simultaneously, with increasing $j$ and $\alpha$  the dimerization
condition mod$( y+2)= y$ fails (Fig. \ref{fig2}),
$f^2_{n+1,n+1}\approx f^2_{n,n}$ and the driving force $F$ vanishes
($x_{max}\rightarrow 0$). The equation (\ref{1otr}) up to second
order terms tends to

\begin{equation}
\frac{1}{2} \ddot{ y} -f_{n+1,n+1}^2  { y}^{''}=
2(f_{n+1,n+1}^2-2f_{n+1,n}^2) y -4f_{n+1,n}^2 y^2. \label{sol}
\end{equation}

If we move to imaginary space $\zeta=i(v\tau-n) $, then the solution
of (\ref{sol}) yields a traveling pulse
\begin{eqnarray}
\lefteqn{ y=  \frac{3\big(f_{n+1,n+1}^2-2f_{n+1,n}^2\big)}{4f_{n+1,n}^2} }   \nonumber \\
& \times  \cosh^{-2}
 \left (\left[{\displaystyle \frac{f_{n+1,n+1}^2-2f_{n+1,n}^2}{2f_{n+1,n+1}^2-v^2}}\right
]^{1/2}(\zeta-\zeta_0)\right ). &  \label{ch}
\end{eqnarray}

From the definition of $\delta_{2n+1}$ we have $x_{2n+2}-x_{2n+1}
\approx
\partial x_{2n+1}/\partial n = 1+\delta_{2n+1}= 2+ y$ and
equation (\ref{ch}) implies a kink-shaped profile,
\begin{eqnarray}
\lefteqn{x_{2n+1}= {\rm const}+ 2n +} \nonumber \\
\lefteqn{\frac{3}{4f_{n+1,n}^2}\Big
[(f_{n+1,n+1}^2-2f_{n+1,n}^2)(2f_{n+1,n+1}^2-v^2)\Big ]^{1/2} } \nonumber \\
& &  \times\tanh \left (\left
 [\frac{f_{n+1,n+1}^2-2f_{n+1,n}^2}{2f_{n+1,n+1}^2-v^2}\right ]^{1/2}(\zeta-\zeta_0)\right
 ),\,
\label{soliton}
\end{eqnarray}
where $v^2< 2f_{n+1,n+1}^2= v_{max}^2$ is the velocity of the
travelling kink profile and
$$L=\frac{1}{2}\left
[\frac{2f_{n+1,n+1}^2-v^2}{f_{n+1,n+1}^2-2f_{n+1,n}^2}\right ]^{1/2}$$
is the kink width.  The time of collision can be estimated from
$v\tau_c= 2L > \left
(2f_{n+1,n+1}^2/\left( f_{n+1,n+1}^2-2f_{n+1,n}^2\right)\right )^{1/2}$ $\sim
\sqrt{\left (  2(j+n_r)/(j-n_r)\right )}$, $j>n_r$. Hence, the
collision time close to the semiclassical limit yields $\tau_{c
}\rightarrow 1/\sqrt {2j} $.

Equation (\ref{phasetr}) with omitted linear terms in the potential
(\ref{Vy}) represents a nonintegrable member of a family of
Klein-Gordon equations, that is,  $\Phi^4$ equation known to exhibit
kink, antikink and a  double kink solitary solutions
\cite{Rajaraman:1982}.
 The solution (\ref{soliton}) represents the
tunneling between two nearest neighbor trajectories. Indeed, for
large $j$ (${f}_{n,n}^2>2f_{n+1,n}^2 $ or $j>n_r$) it gets a form of
a propagating nonlinear solitary excitation, a kink (antikink), in
the imaginary space $y (\zeta-\zeta_0)$, $\zeta=i(v\tau-n)$, where
$\zeta_0=i(v\tau_0-n_0)$ restores the translation symmetry.

The transition to the imaginary space effectively reverts the sign of the
potential, that is, a pseudoparticle turns to the tunneling domain and
 changes its parity  (cf. the mirror
 symmetry of equations (\ref{pairs})). For example,
$\delta_{2n+1}(v_1)\rightarrow \delta_{2n}(v_1)$, $\delta_{2n}(v_2)
\rightarrow \delta_{2n+1}(v_2)$. As a result, the trajectories
(pseudoparticles) interchange their velocities $v_1, v_2$:
$\delta_{2n}(v_2)\rightarrow \delta_{2n}(v_1) $ and
$\delta_{2n+1}(v_1) \rightarrow \delta_{2n+1}(v_2)$  when
transferring from the real to the imaginary space and vice versa.
This scenario can be understood as a series of kink-antikink
collisions related to two subsequent levels.

Thus, the nonlinear phase represents the gCM trajectories as a
kink-antikink chain structure apparent in Figure \ref{fig2}. Such a
solution qualitatively agrees with the results of
authors dealing with similar systems within the gCM level dynamics
approach \cite{Nakamura:1986,Burgdorfer:1992,Gaspard:1989,Zakr:1997}
predicting the ``soliton''-like structures.

The real configuration of the level system under consideration is
described by the discrete label $n$. In recent years more
realistic discrete lattice versions (\ref{deriv}) of Klein-Gordon
equations have been in the focus of interest; they show a rich
variety of behaviour. Besides the resonance energy exchange at
kink-antikink collisions, a  recently discovered aspect of this behaviour is the chaotic
scattering upon certain conditions and fractal behaviour of
velocities after collision strongly dependent on initial conditions
\cite{Anninos:1991,Campbell:1986,Goodman:2007,Vazquez:1992,Dmitriev:2008}.
The kink-antikink collisions in our system might follow this scenario of initiating
the chaotic behaviour and fractal structure at large $j$, Figure \ref{fig3}c.

In the case of $\partial^2 \delta/\partial \tau^2=0$ we have
$p_n={\rm const}$ so that the value of the curvature $K_n=\partial^2
x_n/\partial\tau^2=0$ which corresponds to the peak in Figure \ref{fig5}a
represents the kinks. One has $x_{2n+1}=x_{2n}+v\Delta\tau$ (here
$\Delta\tau$ is the time distance between two kinks) and from
(\ref{pairs}) one finds
\begin{eqnarray}
\delta_{2n+1}+\frac{f_{n,n-1}^2 }{f_{n+1,n}^2}\delta_{2n-1} & \approx &
\nonumber \\
\frac{-2f_{nn}^2+f_{n+1,n}^2+f_{nn-1}^2}{f_{n+1,n}^2} & + &
\frac{2{f}_{nn}^2}{{f}_{n+1
n}^2}\delta_{2n}(1-\delta_{2n})\label{10}
\end{eqnarray}
and
\begin{eqnarray}
\lefteqn{\delta_{2n+1}\approx
\frac{2f_{n+1,n}^2-f_{n+1,n+1}^2-f_{nn}^2}{2f_{n+1,n}^2} + }\label{10'}
\\
& &+ \frac{{f}_{n+1,n+1}^2}{2{f}_{n+1
n}^2}\delta_{2n+2}(1-\delta_{2n+2})+ \frac{{f}_{nn}^2}{2{f}_{n+1
n}^2}\delta_{2n}(1-\delta_{2n})\,.\quad \nonumber
\end{eqnarray}
 Equations
(\ref{10}) and (\ref{10'}) are reminiscent of the logistic equation
$x_{n+1}=Ax_n(1-x_n)$ which implies a known scenario for the
transition to the chaotic region for $A=A_{crit}=3.56994 \dots$ In
our case $A\sim$ $\left({f}_{n+1,n+1}^2+{f}_{n n}^2\right) /2{f
}_{n+1,n}^2$ $\sim \left( n_r+|j-1/2|\right)/n_r$. The ranges of values of $j$ and
$n_r$ considered above give the values of $A \geq 2$ close to $2$ from above,
that is, from the subcritical region. The hallmark of chaos
at the quantum level can be identified  as a pre-chaotic region of
the series of bifurcations of the intermediate glassy phase.

The pre-chaotic behaviour applies to the  medium region of values of
$j$ and $n_r$ between the domains of weak coupling with $j< n_r$
(dimerized pairs of damped oscillators) and strong coupling with
$j\gg n_r$. The description via initial quantum numbers at
$\alpha=0$, $n_r$ and $j$ related to the radial and rotational
degrees of freedom of the unperturbed oscillators, is kept by the
mapping onto the classical gCM gas. For large $j$, $j\gg n_r$, the
short-range correlations are strongly weakened while the long-range
correlations dominate, as well as fluctuating around zero (Fig. \ref{fig2}c).

The kink-antikink chain level structure describes the
special kind of self-avoiding motion of the gCM trajectories in the
integrability parameter $\alpha$ as a pseudotime. Alternatively, one
can also consider the levels as moving along almost straight lines
broken by small intervals  where they undergo "collisions" with
neighbouring levels. They can be clearly visualized as a series of
propagating pulses through the series of gCM pseudoparticles, along
straight lines of positive and negative
velocities. The apparent regularity of this structure allows us to
introduce the idea of a superlattice of excitations in the space of
gCM pseudoparticles. It is this new structure which we call the {\it
kink-train lattice}, which may represent a new {\it quantum
chaotic phase} in certain segments of the JT spectra, created by the
kink-antikink collision mechanism described above.

 A multiplicity of kinks and antikinks (solitary
waves) travelling through gCM particles may be readily observed in Figure 1b of
the another paper on the subject by these authors \cite{Majernikova:2006a}. It is
recognisable in the numerical gCM simulations in Figure \ref{fig2}a, and in the
statistical distribution of velocities of gCM pseudoparticles, Figure
\ref{fig4}. The lack of symmetry of the positive and negative velocities against
zero illustrates the driving of the lattice by a field superimposed
by the broken periodicity, in turn implied by $f_{n+1,n+1}>f_{nn}$
over the whole lattice.

The trajectories forming the {\it kink-antikink train lattice}
 move as almost free particles between the
collisions, with velocities of opposite sign
distributed within two bands (Fig. \ref{fig4}a). The remnants of
oscillations within the gap
demonstrate the coexistence of both oscillating and kink phases. The
long-range order developed at large $\alpha$ stabilises the almost
periodic kink-train lattice (Fig. \ref{fig6}).

It is worth noting that for $j>n_r$ the effective values of the logistic
map constant $A$ can reach the values above the critical $A_c$ for
some $n_r$. This indicates chaotic behaviour of the level
spacings. However, the cumulative spectra (including all $j$) impose
intersections of the trajectories because of the multiple degeneracy
in $j$ which will dominate and imply, for example, the Poisson-like
probability behaviour of the nearest neighbour level spacing
distribution $P(s)$ at $s\rightarrow 0$ \cite{Majernikova:2006b}.
Hence we can conclude that a coexistence of regular and irregular
nonuniversal behaviour is characteristic for the excited
spectra of the present model and in the whole extent of the interaction
parameter.

\subsection{Semiclassical approximation}

One possible way of identifying the quantum chaoticity is to
provide an analogy to the behaviour of the system at a classical
or semiclassical level. The peculiarity of the class of
electron-phonon systems is the marked ambiguity of passing to a
semiclassical approximation from an initial set of Heisenberg
equations implied by Hamiltonian (\ref{1}). The common algorithm,
decoupling products of quantum observables and replacing them by the
classical ones, can be performed in different ways. The problem
of moving to a semiclassical approximation was touched in a
previous paper \cite{Majernikova:2006a}. Nevertheless, the
investigation of trajectories of semiclassical observables
reveals a picture of classically chaotic domains. These at least present
a qualitative correlation to the picture of ``quantum
chaotic'' behaviour outlined above.  Respective equations of motion
are implied by the semiclassical version of the Hamiltonian
(\ref{1}) in the form
\begin{equation}
 H= \frac{1}{2} (P_1^2 + P_2^2 + Q_1^2 + Q_2 ^2) +
\alpha Q_1 z
 -\alpha Q_2 x \,.
\label{class}
\end{equation}

Here, we introduced the coordinates of phonons $Q_i$, $P_i$ and of
the electron $x\equiv \langle \sigma_x\rangle$, $y\equiv
\langle\sigma_y\rangle$, $z=\langle\sigma_z\rangle$. The electron
coordinates satisfy the condition $x^2+y^2+z^2=1$ for the Bloch
sphere. Additionally, besides the conservation of the energy
(\ref{class}), the angular momentum $J=Q_1 P_2-Q_2 P_1 -y$ is a
constant of motion. As a result, the system has four degrees of freedom.

\begin{figure*}[htb]
\includegraphics[width=0.6\hsize]{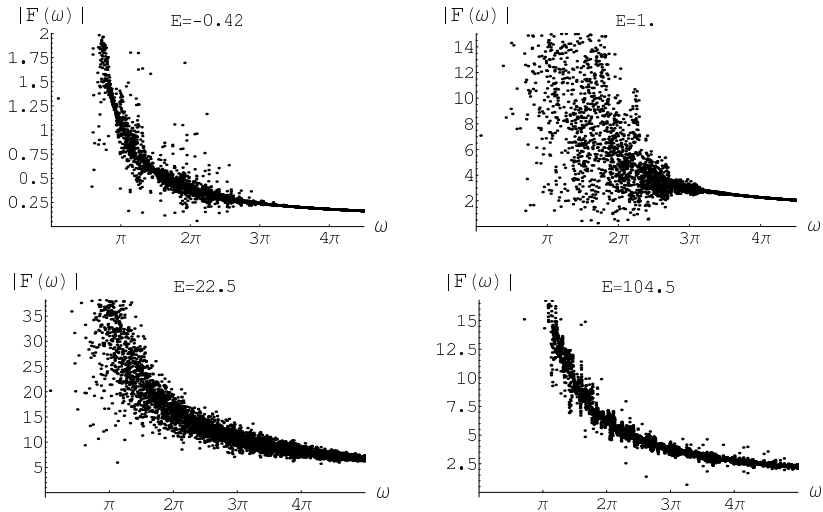}
\caption{ Frequency spectrum of the $Q_1$ trajectory in
semiclassical approximation for different values of total energy }
\label{fig8}
\end{figure*}

For illustrative purposes, in Figure \ref{fig8} we present  the Fourier spectrum of the
semiclassical trajectory $Q_1(t)$ (a picture for $Q_2$
looks similar) for different values of the energy $E$ of a system.
The middle part of the energy interval (corresponding approximately
 to the intermediate part of the quantum spectrum with the strongest
mixing of three effective potential wells) shows  the marked
domain of classical chaos. Meanwhile, the motion of the classical
trajectories at low and high energies is noticeably regular.

\section{Conclusion}
\label{concl} The dynamics of the excited levels of the E$\otimes$e
Jahn-Teller model, mapped onto the generalized Calogero-Moser gas of
interacting pseudoparticles and developing in pseudotime, shows a
complex interplay between the nonlinearity and fluctuations of the
dynamical degrees of freedom. Numerical results of Section \ref{num} on
lattice dynamics and the probabilistic characteristics were
interpreted, at least qualitatively, on the basis of the approximate
analytical results of Section \ref{analyt}.

In the intermediate range, $j\sim n_r$, the gCM dynamics implies a
maximum degree of irregularity caused by the interference of three
branches of the pseudocharges with growing range of
interaction. We have found that the maximum mixing, and hence the
broad maximum of stochasticity (irregularity) illustrated by the
entropy of curvatures of the gCM trajectories (accelerations), is
related to the new intermediate {\it glassy phase of nucleation}.
This is thought to represent
the growing kink-antikink chain structure. This scenario can be
thought of for each level as an analogous to a series of local
first order quantum phase transitions. In the intermediate
droplet glassy phase, the phase of damped (anharmonic)
oscillations dominating at $j<n_r$ and the lattice of interacting
kink-antikink chains of the gCM trajectories dominating for $j>n_r$ coexist.
As an alternative picture of this chain structure, we have recognised
the {\it kink-train lattice  phase} of
 the levels. These move along straight lines broken by small intervals, where they undergo
``collisions'' (avoidings) with neighbour levels. They form a {\it superlattice of
excitations} in the space of gCM pseudoparticles which we believe to
represent a {\it new quantum chaotic phase}. The trajectories forming the
 kink train lattice of the kinks and antikinks move almost as free particles between collisions, with velocities of opposite sign distributed within two bands  (Fig. \ref{fig4}a).  The long-range
order developed at large $\alpha$ stabilises the almost periodic
kink-train lattice (Fig. \ref{fig6}). It is worth remembering that throughout this
work we used the rotational quantum number $j$ as a parameter. Real
spectral characteristics are cumulative ones including the
contributions of all $j$'s, where the multiple degeneracy in $j$
(level crossings) brings in components of regularity. Hence,
real statistical properties present the mixing (coexistence)
of contributions with prevailing higher degree of regularity or of
those with more irregular nature. These are, however, highly
nonuniversal.


\vspace{1em}

  The support of the project No. 202/06/0396 by the Grant Agency of the Czech Republic
is greatly acknowledged as well as the support by the Grant Agency
VEGA, Bratislava of the project No. 2/6073/26.

%

\end{document}